\documentclass[10pt,twocolumn,english,aps,manuscript,aps,manuscript,aps,showpacs,showkeys,superscriptaddress,citeautoscript]{revtex4}
\usepackage[T1]{fontenc}
\usepackage[latin9]{inputenc}
\usepackage[active]{srcltx}
\usepackage{color}
\usepackage{float}
\usepackage{multirow}
\usepackage{amsmath}
\usepackage{amssymb}
\usepackage{graphicx}
\usepackage{wasysym}

\makeatletter

\providecommand{\tabularnewline}{\\}

\@ifundefined{textcolor}{}
{%
 \definecolor{BLACK}{gray}{0}
 \definecolor{WHITE}{gray}{1}
 \definecolor{RED}{rgb}{1,0,0}
 \definecolor{GREEN}{rgb}{0,1,0}
 \definecolor{BLUE}{rgb}{0,0,1}
 \definecolor{CYAN}{cmyk}{1,0,0,0}
 \definecolor{MAGENTA}{cmyk}{0,1,0,0}
 \definecolor{YELLOW}{cmyk}{0,0,1,0}
}

\usepackage{babel}

\makeatother

\usepackage{babel}
\begin{document}
\title{Peculiarities in pseudo-transitions of a mixed spin-$(1/2,1)$ Ising-Heisenberg
double-tetrahedral chain in an external magnetic field}
\author{Onofre Rojas}
\affiliation{Departamento de Física, Universidade Federal de Lavras, CP 3037, 37200-000,
Lavras-MG, Brazil}
\author{Jozef Stre\v{c}ka}
\affiliation{Department of Theoretical Physics and Astrophysics, Faculty of Science,
P. J. Šafárik University, Park Angelinum 9, 040 01 Košice, Slovakia}
\author{Oleg Derzhko}
\affiliation{Institute for Condensed Matter Physics, National Academy of Sciences
of Ukraine, Svientsitskii Str. 1, 79011 L'viv, Ukraine}
\author{S. M. de Souza}
\affiliation{Departamento de Física, Universidade Federal de Lavras, CP 3037, 37200-000,
Lavras-MG, Brazil}
\begin{abstract}
Recently, it has been rigorously verified that several one-dimensional
(1D) spin models may exhibit a peculiar pseudo-transition accompanied
with anomalous response of thermodynamic quantities in a close vicinity
of pseudo-critical temperature. In the present work we will introduce
and exactly solve a mixed spin-(1/2,1) Ising-Heisenberg double-tetrahedral
chain in an external magnetic field as another particular example
of 1D lattice-statistical model with short-range interactions that
displays a pseudo-transition of this type. The investigated model
exhibits at zero temperature three ferrimagnetic phases, three frustrated
phases, and one saturated paramagnetic phase. The ground-state phase
diagram involves five unusual interfaces (phase boundaries), at which
the residual entropy per site equals to a larger entropy of one of
two coexisting phases. Four such interfaces are between a non-degenerate
ferrimagnetic phase and a macroscopically degenerate frustrated phase,
while one interface is between two non-degenerate ferrimagnetic phases.
Though thermal excitations typically destroy all fingerprints of zero-temperature
phase transitions of 1D lattice-statistical models with short-range
forces, the mixed spin-(1/2,1) Ising-Heisenberg double-tetrahedral
chain is quite robust with respect to thermal excitations and it displays
peculiar pseudo-transitions close to all five aforementioned interfaces. 
\end{abstract}
\maketitle

\section{Introduction}

\label{Sec1}

There are a few paradigmatic examples of one-dimensional (1D) lattice-statistical
models with short-range couplings, which exhibit a discontinuous (first-order)
phase transition at finite temperature. Perhaps the most famous example
is 1D KDP model of hydrogen-bonded ferroelectrics invented by Nagle
\citep{nagle}, which displays a discontinuous phase transition between
the ferroelectric and paraelectric phases due to assignment of an
infinite energy to all ionized configurations. Another particular
example of this type is the Kittel model \citep{kittel} defined through
a finite transfer matrix, which involves a constraint on zipper corresponding
to an infinite potential being responsible for a non-analyticity of
the free energy. Owing to a singular character of the potential, the
Kittel model also exhibits a first-order phase transition. The next
paradigmatic example is the 1D solid-on-solid model considered by
Chui and Weeks \citep{chui}, which is exactly solvable in spite of
an infinite dimension of its transfer matrix. By imposing suitable
pinning potential the 1D solid-on-solid model may also display a roughening
phase transition of first order \citep{chui}. Furthermore, Dauxois
and Peyrard \citep{dauxois} have examined another 1D lattice-statistical
model with an infinite dimension of the transfer matrix, which exhibits
a phase transition at finite temperature. Last but not least, Sarkanych
\textit{et al}. \citep{sarkanych} proposed 1D Potts model with so-called
invisible states and short-range couplings. It could be thus concluded
that all five aforementioned 1D lattice-statistical models break the
Perron-Frobenius theorem, because some off-diagonal transfer-matrix
elements become null and the free energy may consequently become non-analytic
at a certain critical temperature.

Van Hove \citep{Hove} proposed a theorem that proves absence of a
phase transition in 1D lattice-statistical models with short-range
couplings. Later, Cuesta and Sanchez \citep{cuesta} generalized the
non-existence theorem for a phase transition at finite temperatures.
Surely, this is not yet the most general non-existence theorem, because
mixed-particle chains or more general external fields fall beyond
the scope of this theorem.

The term "pseudo-transition" and "quasi-phase" was introduced
by Timonin \citep{Timonin} in 2011 when studying the spin-ice model
in a field. These terms refer to a sudden change in first derivatives
and vigorous peaks in second derivatives of the free energy although
these marked signatures are not in reality true discontinuities and
divergences, respectively. Note furthermore that the pseudo-transitions
do not violate the Perron-Frobenius theorem, because the free energy
is always analytic. A common feature of the pseudo-transitions is
that some off-diagonal transfer-matrix elements (Boltzmann factors)
become very small (almost zero), since very high albeit finite energy
is assigned to the corresponding states. 

Obvious fingerprints of pseudo-transitions were recently found in
several 1D spin or spin-electron models. For instance, the pseudo-transitions
were detected in the spin-1/2 Ising-Heisenberg diamond chain \citep{Isaac2,Isaac},
two-leg ladder \citep{on-strk}, as well as triangular tube \citep{strk-cav}.
Similarly, the emergence of pseudo-transitions was verified in the
spin-1/2 Ising diamond chain \citep{csmag} and the coupled spin-electron
double-tetrahedral chain \citep{Galisova,galisova17,galisova18}.
In general, the first derivatives of the free energy such as entropy,
internal energy or magnetization show a steep change around pseudo-critical
temperature. This feature is similar to the first-order phase transition,
but all thermodynamic response functions are in fact continuous. Contrary
to this, second derivatives of the free energy such as specific heat
and magnetic susceptibility resemble typical behavior of a second-order
phase transition at a finite temperature. Therefore, this peculiar
pseudo-critical behavior drew attention to a more comprehensive study
of this phenomenon aimed at elucidating all its essential features
\citep{pseudo,ph-bd,tk}. Recently, a further attention has been paid
to uncover the mechanism triggering pseudo-transitions based on a
rigorous analysis of the correlation function \citep{Isaac} and pseudo-critical
exponents \citep{expo}.

The goal of the present study is to investigate a mixed spin-(1/2,1)
Ising-Heisenberg tetrahedral chain in an external magnetic field,
which has a pretty rich ground-state phase diagram and exhibits a
number of finite-temperature pseudo-transitions close to some inter-phase
boundaries.There are some 3D compounds in which, when we consider
one columnar stripe, we could observe a double tetrahedral chain structure.
Such as cobalt oxide $\mathrm{RBaCo_{4}}\mathrm{O}_{7}$, where $\mathrm{R}$
denotes a rare earth atom, which has a swedenborgite lattice structure\citep{fritz}.
Another compound with a similar structure could be the salt with 3D
corrugated packing frustrated spin \citep{otsuka} of $\mathrm{C_{60}^{\bullet-}}$
in ($\mathrm{MDABCO^{+})(C_{60}^{\bullet-})}$ {[}$\mathrm{MDABCO^{+}}=N$-methyldiazabicyclooctanium
cation and $\mathrm{C}_{60}^{\bullet-}$ radical anions{]}, a stripe
of this salt can be viewed also as a double-tetrahedral chain.

This article is organized as follows. In Sec.~\ref{Sec2} we consider
and exactly solve the mixed spin-(1/2,1) Ising-Heisenberg tetrahedral
chain in a magnetic field. Thermodynamics in a close vicinity of the
pseudo-transition is examined in Sec.~\ref{Sec3}, where an influence
of the residual entropy upon basic thermodynamic quantities is investigated
in detail. Finally, several concluding remarks are presented in Sec.~\ref{Sec4}.

\section{Mixed spin-($1/2,1)$ Ising-Heisenberg double-tetrahedral chain}

\label{Sec2}

The coupled spin-electron model on a double-tetrahedral chain \citep{Galisova,galisova17,galisova18},
which involves localized Ising spins at nodal lattice sites and mobile
electrons delocalized over triangular plaquettes, represents a prominent
example of 1D lattice-statistical model mimicking a temperature-driven
phase transition \citep{Galisova}. However, earlier investigations
of the analogous spin-1/2 Heisenberg \citep{mambrini,roj-alc,Maksymenko2011}
and Ising-Heisenberg \citep{vadim-1,Vadim-2} models on a double-tetrahedral
chain did not verify anomalous thermodynamic response closely related
to a pseudo-transition until the latter Ising-Heisenberg model was
revisited and more thoroughly studied \citep{ph-bd}.

\begin{figure}[H]
\centering{} \includegraphics[scale=0.8]{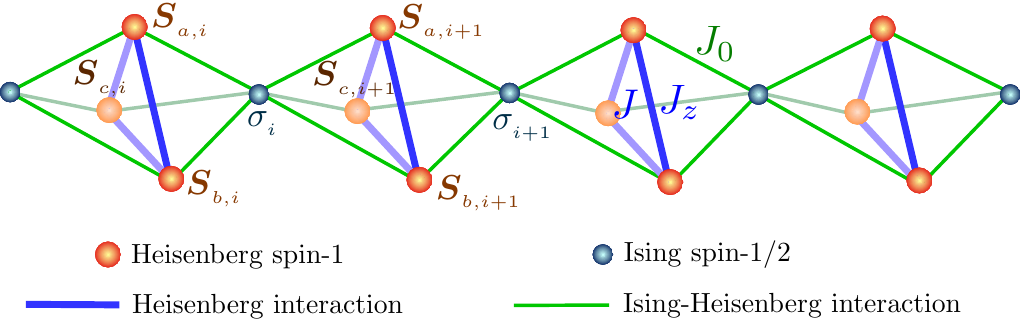} \caption{\label{fig:Sch-trd-ch} A schematic representation of the mixed spin-(1/2,1)
Ising-Heisenberg double-tetrahedral chain. Small balls correspond
to the Ising spins $\sigma_{i}$ and large balls correspond to the
Heisenberg spins $\boldsymbol{S}_{\gamma,i}(\gamma=a,b,c)$.}
\end{figure}

In the present work we will examine in particular the mixed spin-($1/2,1$)
Ising-Heisenberg double-tetrahedral chain, which is schematically
depicted in Fig.~\ref{fig:Sch-trd-ch} and defined through the following
Hamiltonian 
\begin{equation}
H=\sum_{i=1}^{N}H_{i},
\end{equation}
with 
\begin{alignat}{1}
H_{i} & =-[J(\boldsymbol{S}_{b,i},\boldsymbol{S}_{c,i})_{z}+J(\boldsymbol{S}_{c,i},\boldsymbol{S}_{a,i})_{z}+J(\boldsymbol{S}_{a,i},\boldsymbol{S}_{b,i})_{z}]\nonumber \\
 & \!-\!\left(S_{a,i}^{z}\!+\!S_{b,i}^{z}\!+\!S_{c,i}^{z}\right)\!\left[h_{z}\!+\!J_{0}(\sigma_{i}\!+\!\sigma_{i+1})\right]\!-\!\tfrac{h}{2}\!\left(\sigma_{i}\!+\!\sigma_{i+1}\right)\!.\label{eq:H-tetra}
\end{alignat}
In above, $S_{\gamma,i}^{\alpha}$ ($\alpha=\{x,y,z\}$, $\gamma=\{a,b,c\}$)
denote the spin-1 Heisenberg atoms, $\sigma_{i}=\pm\frac{1}{2}$ denotes
the Ising spin, and $J(\boldsymbol{S}_{\gamma,i},\boldsymbol{S}_{\delta,i})_{z}=JS_{\gamma,i}^{x}S_{\delta,i}^{x}+JS_{\gamma,i}^{y}S_{\delta,i}^{y}+J_{z}S_{\gamma,i}^{z}S_{\delta,i}^{z}$.
The Hamiltonian (\ref{eq:H-tetra}) is written as a sum of cell Hamiltonians
$H_{i}$, which correspond to spin clusters with the geometric shape
of two face-sharing tetrahedra (i.e., trigonal bipyramid).

The overall Hilbert space of the mixed spin-($1/2,1$) Ising-Heisenberg
double-tetrahedral chain splits into several disjoint (orthogonal)
subspaces, because the Hamiltonians $H_{i}$ from different unit cells
commute with each other. The Hilbert subspace corresponding to the
spin-1 Heisenberg triangle from the $i$-th unit cell is given by
the Hamiltonian matrix of dimension $27\times27$ and it can be further
split into several smaller block-diagonal matrices depending on the
$z$-component of the total spin: for $S_{t}^{z}=0$ one has one $7\times7$
block matrix, for $|S_{t}^{z}|=1$ two $6\times6$ matrices, for $|S_{t}^{z}|=2$
two $3\times3$ matrices, and for $|S_{t}^{z}|=3$ two $1\times1$
matrices. All eigenvalues and eigenvectors of spin-1 Heisenberg triangle
Hamiltonian are listed in Table~\ref{tab1}. The
first column stands for the eigenvalues of the $S_{t}^{z}$ operator,
while the counter $k$ is used just to distinguish the states with
same eigenvalues and the respective state degeneracy $g_{k}$ in fourth
column. With the help of eigenvalues and eigenvectors of the spin-1
Heisenberg triangle reported in Table \ref{tab1} one can express
the full energy spectrum per $H_{i}$ unit cell of the mixed spin-($1/2,1$)
Ising-Heisenberg double-tetrahedral chain as follows 
\begin{equation}
\varepsilon_{k}\left(\sigma_{i},\sigma_{i+1}\right)=\epsilon_{k}-\left(J_{0}S_{t}^{z}+\frac{h}{2}\right)\left(\sigma_{i}+\sigma_{i+1}\right).\label{eq:E-spct}
\end{equation}
Here, $\epsilon_{k}$ marks the respective eigenvalue of the spin-1
Heisenberg triangle listed in Table \ref{tab1}.

\begin{widetext}

\begin{table}[h]
\caption{\label{tab1} Full spectrum of the spin-1 Heisenberg triangle specified
according to the respective eigenvalue, state degeneracy, and eigenvector.
The eigenstates are grouped according to the $z$-component of the
total spin $S_{t}^{z}=S_{a}^{z}+S_{b}^{z}+S_{c}^{z}$. The
first column stands for the eigenvalues of the $S_{t}^{z}$ operator,
and the second column is just to distinguish the eigenvector with
the same $S_{t}^{z}$. The definition of mixing angles: $\cot\left(2\phi_{1}\right)=\frac{J_{z}-J}{2J}$,
$\cot\left(2\phi_{2}\right)=\frac{J_{z}+2J}{4J}$, and $\cot\left(2\phi_{3}\right)=\frac{J_{z}-2J}{2\sqrt{6}J}$.}
\vspace{3mm}
\begin{tabular}{|c|l|l|l|l|}
\hline 
$|S_{t}^{z}|$  & $k$  & Energy ($\epsilon_{k}$)  & $g_{k}$  & State\tabularnewline
\hline 
\hline 
\multirow{5}{*}{$0$} & $0$  & $\mbox{\small\ensuremath{J+J_{z}}}$  & $2$  & $\mbox{\scriptsize|0,0\ensuremath{\rangle=}\ensuremath{\begin{cases}
\tfrac{1}{2}\left(\left|\substack{1\\
0\\
-1
}
\right\rangle -\left|\substack{0\\
1\\
-1
}
\right\rangle -\left|\substack{0\\
-1\\
1
}
\right\rangle +\left|\substack{-1\\
0\\
1
}
\right\rangle \right)\\
\tfrac{\sqrt{3}}{6}\left(2\left|\substack{1\\
-1\\
0
}
\right\rangle -\left|\substack{1\\
0\\
-1
}
\right\rangle -\left|\substack{0\\
1\\
-1
}
\right\rangle -\left|\substack{0\\
-1\\
1
}
\right\rangle -\left|\substack{-1\\
0\\
1
}
\right\rangle +2\left|\substack{-1\\
1\\
0
}
\right\rangle \right)
\end{cases}}}$\tabularnewline
\cline{2-5} \cline{3-5} \cline{4-5} \cline{5-5} 
 & $1$  & $\mbox{\small\ensuremath{J\sqrt{6}\cot\phi_{3}}}$  & $1$  & $\mbox{\scriptsize|0,1\ensuremath{\rangle}=\ensuremath{\tfrac{\sqrt{6}}{6}\cos\phi_{3}\left(\left|\substack{1\\
0\\
-1
}
\right\rangle +\left|\substack{1\\
-1\\
0
}
\right\rangle +\left|\substack{0\\
1\\
-1
}
\right\rangle +\left|\substack{0\\
-1\\
1
}
\right\rangle +\left|\substack{-1\\
1\\
0
}
\right\rangle +\left|\substack{-1\\
0\\
1
}
\right\rangle \right)}-\ensuremath{\sin}\ensuremath{\phi_{3}\left|\substack{0\\
0\\
0
}
\right\rangle }}$\tabularnewline
\cline{2-5} \cline{3-5} \cline{4-5} \cline{5-5} 
 & $2$  & $\mbox{\small\ensuremath{-J\sqrt{6}\tan\phi_{3}}}$  & $1$  & $\mbox{\scriptsize\ensuremath{|0,2\rangle=}\ensuremath{\tfrac{\sqrt{6}}{6}\sin\phi_{3}\left(\left|\substack{1\\
0\\
-1
}
\right\rangle +\left|\substack{1\\
-1\\
0
}
\right\rangle +\left|\substack{0\\
1\\
-1
}
\right\rangle +\left|\substack{0\\
-1\\
1
}
\right\rangle +\left|\substack{-1\\
1\\
0
}
\right\rangle +\left|\substack{-1\\
0\\
1
}
\right\rangle \right)}+\ensuremath{\cos\phi_{3}\left|\substack{0\\
0\\
0
}
\right\rangle }}$\tabularnewline
\cline{2-5} \cline{3-5} \cline{4-5} \cline{5-5} 
 & $3$  & $\mbox{\small\ensuremath{-J+J_{z}}}$  & $2$  & $\mbox{\scriptsize\ensuremath{|0,3\rangle=\left\{ \hspace{-0.2cm}\begin{array}{l}
\tfrac{1}{2}\left(-\left|\substack{1\\
0\\
-1
}
\right\rangle -\left|\substack{0\\
1\\
-1
}
\right\rangle +\left|\substack{0\\
-1\\
1
}
\right\rangle +\left|\substack{-1\\
0\\
1
}
\right\rangle \right)\\
\tfrac{\sqrt{3}}{6}\left(-2\left|\substack{1\\
-1\\
0
}
\right\rangle -\left|\substack{1\\
0\\
-1
}
\right\rangle -\left|\substack{0\\
-1\\
1
}
\right\rangle +\left|\substack{0\\
1\\
-1
}
\right\rangle +\left|\substack{-1\\
0\\
1
}
\right\rangle +2\left|\substack{-1\\
1\\
0
}
\right\rangle \right)
\end{array}\right.}}$\tabularnewline
\cline{2-5} \cline{3-5} \cline{4-5} \cline{5-5} 
 & $4$  & $\mbox{\small\ensuremath{2J+J_{z}}}$  & $1$  & $\mbox{\scriptsize\ensuremath{|0,4\rangle=}\ensuremath{\tfrac{\sqrt{6}}{6}\left(-\left|\substack{1\\
0\\
-1
}
\right\rangle +\left|\substack{1\\
-1\\
0
}
\right\rangle +\left|\substack{0\\
1\\
-1
}
\right\rangle -\left|\substack{0\\
-1\\
1
}
\right\rangle -\left|\substack{-1\\
1\\
0
}
\right\rangle +\left|\substack{-1\\
0\\
1
}
\right\rangle \right)}}$\tabularnewline
\hline 
\multirow{4}{*}{$1$} & \multicolumn{1}{l|}{$\begin{array}{c}
5\\
6
\end{array}$} & $\mbox{\small\ensuremath{-2J\left(1-\cot\phi_{2}\right)\pm h_{z}}}$  & $1$  & $\mbox{\scriptsize\ensuremath{|\pm1,0\rangle=}\ensuremath{\tfrac{\sqrt{3}}{3}\left[\cos\phi_{2}\left(\left|\substack{\pm1\\
\pm1\\
\mp1
}
\right\rangle +\left|\substack{\pm1\\
\mp1\\
\pm1
}
\right\rangle +\left|\substack{\mp1\\
\pm1\\
\pm1
}
\right\rangle \right)-\sin\phi_{2}\left(\left|\substack{\pm1\\
0\\
0
}
\right\rangle +\left|\substack{0\\
\pm1\\
0
}
\right\rangle +\left|\substack{0\\
0\\
\pm1
}
\right\rangle \right)\right]}}$\tabularnewline
\cline{2-5} \cline{3-5} \cline{4-5} \cline{5-5} 
 & $\begin{array}{c}
7\\
8
\end{array}$  & $-2J\left(1+\tan\phi_{2}\right)\pm h_{z}$  & $1$  & $\mbox{\scriptsize\ensuremath{|\pm1,1\rangle=}\ensuremath{\tfrac{\sqrt{3}}{3}\left[\sin\phi_{2}\left(\left|\substack{\pm1\\
\pm1\\
\mp1
}
\right\rangle +\left|\substack{\pm1\\
\mp1\\
\pm1
}
\right\rangle +\left|\substack{\mp1\\
\pm1\\
\pm1
}
\right\rangle \right)+\cos\phi_{2}\left(\left|\substack{\pm1\\
0\\
0
}
\right\rangle +\left|\substack{0\\
\pm1\\
0
}
\right\rangle +\left|\substack{0\\
0\\
\pm1
}
\right\rangle \right)\right]}}$\tabularnewline
\cline{2-5} \cline{3-5} \cline{4-5} \cline{5-5} 
 & $\begin{array}{c}
9\\
10
\end{array}$  & $\mbox{\small\ensuremath{J\left(1+\cot\phi_{1}\right)\mp h_{z}}}$  & $2$  & $\mbox{\scriptsize\ensuremath{|\pm1,2\rangle=\left\{ \hspace{-0.2cm}\begin{array}{l}
\tfrac{\sqrt{2}}{2}\left[\sin\phi_{1}\left(\left|\substack{0\\
0\\
\pm1
}
\right\rangle -\left|\substack{0\\
\pm1\\
0
}
\right\rangle \right)+\cos\phi_{1}\left(\left|\substack{\pm1\\
\pm1\\
\mp1
}
\right\rangle -\left|\substack{\pm1\\
\mp1\\
\pm1
}
\right\rangle \right)\right]\\
\tfrac{\sqrt{6}}{6}\left[\cos\phi_{1}\left(2\left|\substack{\mp1\\
\pm1\\
\pm1
}
\right\rangle -\left|\substack{\pm1\\
\mp1\\
\pm1
}
\right\rangle -\left|\substack{\pm1\\
\pm1\\
\mp1
}
\right\rangle \right)+\sin\phi_{1}\left(2\left|\substack{\pm1\\
0\\
0
}
\right\rangle -\left|\substack{0\\
\pm1\\
0
}
\right\rangle -\left|\substack{0\\
0\\
\pm1
}
\right\rangle \right)\right]
\end{array}\right.}}$\tabularnewline
\cline{2-5} \cline{3-5} \cline{4-5} \cline{5-5} 
 & $\begin{array}{c}
11\\
12
\end{array}$  & $\mbox{\small\ensuremath{J\left(1-\tan\phi_{1}\right)\mp h_{z}}}$  & $2$  & $\mbox{\scriptsize\ensuremath{|\pm1,3\rangle=\left\{ \hspace{-0.2cm}\begin{array}{l}
\tfrac{\sqrt{2}}{2}\left[\cos\phi_{1}\left(\left|\substack{\pm1\\
0\\
0
}
\right\rangle -\left|\substack{0\\
\pm1\\
0
}
\right\rangle \right)+\sin\phi_{1}\left(\left|\substack{\pm1\\
\mp1\\
\pm1
}
\right\rangle -\left|\substack{\mp1\\
\pm1\\
\pm1
}
\right\rangle \right)\right]\\
\tfrac{\sqrt{6}}{6}\left[\sin\phi_{1}\left(2\left|\substack{\pm1\\
\pm1\\
\mp1
}
\right\rangle -\left|\substack{\pm1\\
\mp1\\
\pm1
}
\right\rangle -\left|\substack{\mp1\\
\pm1\\
\pm1
}
\right\rangle \right)-\cos\phi_{1}\left(2\left|\substack{0\\
0\\
\pm1
}
\right\rangle -\left|\substack{0\\
\pm1\\
0
}
\right\rangle -\left|\substack{\pm1\\
0\\
0
}
\right\rangle \right)\right]
\end{array}\right.}}$\tabularnewline
\hline 
\multirow{2}{*}{$2$} & $\begin{array}{c}
13\\
14
\end{array}$  & $\mbox{\small\ensuremath{J-J_{z}\mp2h_{z}}}$  & $2$  & $\mbox{\scriptsize\ensuremath{|\pm2,0\rangle=\begin{cases}
\tfrac{\sqrt{2}}{2}\left(\left|\substack{\pm1\\
\pm1\\
0
}
\right\rangle -\left|\substack{0\\
\pm1\\
\pm1
}
\right\rangle \right)\\
\tfrac{\sqrt{6}}{6}\left(\left|\substack{\pm1\\
\pm1\\
0
}
\right\rangle -2\left|\substack{\pm1\\
0\\
\pm1
}
\right\rangle +\left|\substack{0\\
\pm1\\
\pm1
}
\right\rangle \right)
\end{cases}}}$\tabularnewline
\cline{2-5} \cline{3-5} \cline{4-5} \cline{5-5} 
 & $\begin{array}{c}
15\\
16
\end{array}$  & $\mbox{\small\ensuremath{-2J-J_{z}\mp2h_{z}}}$  & $1$  & $\mbox{\scriptsize\ensuremath{|\pm2,1\rangle=}\ensuremath{\tfrac{\sqrt{3}}{3}\left(\left|\substack{\pm1\\
\pm1\\
0
}
\right\rangle +\left|\substack{\pm1\\
0\\
\pm1
}
\right\rangle +\left|\substack{0\\
\pm1\\
\pm1
}
\right\rangle \right)}}$\tabularnewline
\hline 
$3$  & $\begin{array}{c}
17\\
18
\end{array}$  & $\mbox{\small\ensuremath{-3J_{z}\mp3h_{z}}}$  & $1$  & $\mbox{\scriptsize\ensuremath{|\pm3,0\rangle=}\ensuremath{\left|\substack{\pm1\\
\pm1\\
\pm1
}
\right\rangle }}$\tabularnewline
\hline 
\end{tabular}
\end{table}

\end{widetext}

\subsection{Ground-state phase diagram}

\begin{figure}[h]
\begin{centering}
\includegraphics[scale=0.45]{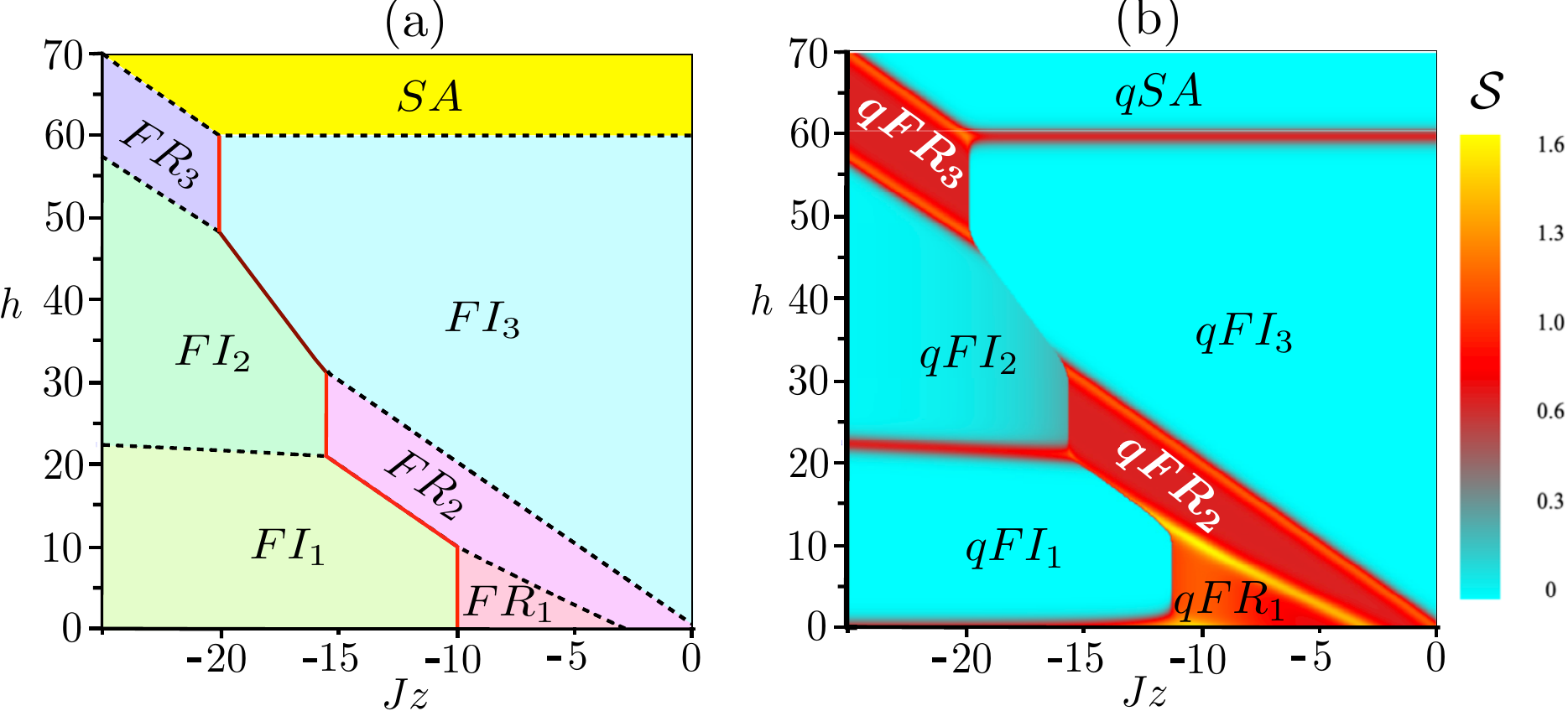} 
\par\end{centering}
\caption{\label{fig:Zero-temperature} (a) Ground-state phase diagram in the
$J_{z}-h$ plane by assuming the fixed parameters $J=-10$, $J_{0}=-10$,
and $h_{z}=h$; (b) Density-plot of entropy in the $J_{z}-h$ plane
for the same set of parameters as in (a) at $T=0.4$.}
\end{figure}

The ground-state phase diagram shown in Fig.~\ref{fig:Zero-temperature}(a)
totally involves seven phases specified below. First, the saturated
paramagnetic phase ($SA$) has according to Eq.~\eqref{eq:E-spct}
the following energy per unit cell 
\begin{equation}
E_{SA}=-3J_{0}-3J_{z}-3h_{z}-\tfrac{1}{2}h,
\end{equation}
which corresponds to the eigenstate defined through the eigenvector
$\left|3,0\right\rangle _{i}$ specified in Table~\ref{tab1} 
\begin{equation}
|SA\rangle=\prod_{i=1}^{N}\left|3,0\right\rangle _{i}|+\rangle_{i}.\label{Trt-SA}
\end{equation}
Obviously, both Ising spin magnetization per unit cell ($m_{I}=\frac{1}{2}$)
and Heisenberg spin magnetization per unit cell ($m_{H}=3$) are fully
polarized, and total magnetization per unit cell attains the following
value $m_{t}=m_{I}+m_{H}=\tfrac{7}{2}$. 

The ground-state phase diagram shown in Fig.~\ref{fig:Zero-temperature}(a)
also displays three different ferrimagnetic ($FI$) phases. The ground-state
energy of the first ferrimagnetic phase $FI_{1}$ reads 
\begin{alignat}{1}
E_{FI_{1}}= & 2J+J_{z}-\tfrac{1}{2}h,
\end{alignat}
whereas its corresponding eigenvector is given by 
\begin{equation}
|FI_{1}\rangle=\prod_{i=1}^{N}\left|0,4\right\rangle _{i}|+\rangle_{i}\label{Trt-FI1}
\end{equation}
with the eigenvector $\left|0,4\right\rangle _{i}$ defined in Table~\ref{tab1}.
In the first ferrimagnetic phase $FI_{1}$ the Ising spin magnetization
is $m_{I}=\frac{1}{2}$, the Heisenberg spin magnetization equals
zero $m_{H}=0$, and the total magnetization thus becomes $m_{t}=\tfrac{1}{2}$.

The ground-state energy for the second ferrimagnetic phase $FI_{2}$
can be expressed as 
\begin{alignat}{1}
E_{FI_{2}}= & -J_{0}-2J\left(1-\cot\phi_{2}\right)-h_{z}-\tfrac{1}{2}h,
\end{alignat}
where $\cot\left(2\phi_{2}\right)=\frac{J_{z}+2J}{4J}$ with $-\frac{\pi}{4}<\phi_{2}<\frac{\pi}{4}$.
The corresponding eigenvector reads 
\begin{equation}
|FI_{2}\rangle=\prod_{i=1}^{N}\left|1,1\right\rangle _{i}|+\rangle_{i}\label{Trt-FI2}
\end{equation}
with the eigenvector $\left|1,1\right\rangle _{i}$ defined in Table~\ref{tab1}.
The Ising spin magnetization in the second ferrimagnetic phase $FI_{2}$
becomes $m_{I}=\frac{1}{2}$, the Heisenberg spin magnetization is
$m_{H}=1$, and the total magnetization is $m_{t}=\tfrac{3}{2}$.

The ground-state energy for the third ferrimagnetic phase $FI_{3}$
is given by 
\begin{alignat}{1}
E_{FI_{3}}= & 3J_{0}-3J_{z}-3h_{z}+\tfrac{1}{2}h,
\end{alignat}
whereas its corresponding eigenvector reads 
\begin{equation}
|FI_{3}\rangle=\prod_{i=1}^{N}\left|3,0\right\rangle _{i}|-\rangle_{i}\label{Trt-FI}
\end{equation}
with the eigenvector $\left|3,0\right\rangle _{i}$ being defined
in Table~\ref{tab1}. Analogously to the previous case, the Ising
spin magnetization is given by $m_{I}=-\frac{1}{2}$, the Heisenberg
spin magnetization equals to $m_{H}=3$, and the total magnetization
is $m_{t}=\tfrac{5}{2}$. It should be pointed out that the saturated
paramagnetic phase as well as all three ferrimagnetic phases are non-degenerate,
which means that there is no residual entropy $\mathcal{S}=0$ at
zero temperature within those ground states.

However, the ground state of the mixed spin-($1/2,1$) Ising-Heisenberg
double-tetrahedral chain may be one of three frustrated ($FR$) phases
with a nonzero residual entropy. The ground-state energy of the first
frustrated phase $FR_{1}$ is given by 
\begin{alignat}{1}
E_{_{FR_{1}}}= & J_{0}-J\left(1+\cot\phi_{1}\right)-h_{z}+\tfrac{1}{2}h,
\end{alignat}
where $\cot\left(2\phi_{1}\right)=\frac{J_{z}-J}{2J}$ with $-\frac{\pi}{4}<\phi_{1}<\frac{\pi}{4}$.
The corresponding ground-state eigenvector reads as follows 
\begin{equation}
|FR_{1}\rangle=\prod_{i=1}^{N}\left|1,3\right\rangle _{i}|-\rangle_{i},\label{Trt-FR1}
\end{equation}
where two-fold degenerate eigenstate $\left|1,3\right\rangle _{i}$
is specified in Table~\ref{tab1}. Owing to this fact, the frustrated
phase $FR_{1}$ is macroscopically degenerate with the residual entropy
$\mathcal{S}=\ln(2)$ per unit cell when the entropy is measured in
units of the Boltzmann constant $k_{B}$. Note that the Ising spin
magnetization is being $m_{I}=-\frac{1}{2}$, the Heisenberg spin
magnetization is $m_{H}=1$, and the total magnetization becomes $m_{t}=\tfrac{1}{2}$.

The ground-state energy of the second frustrated phase $FR_{2}$ can
be expressed as follows 
\begin{alignat}{1}
E_{_{FR_{2}}}=2J_{0}+ & J-J_{z}-2h_{z}+\tfrac{1}{2}h
\end{alignat}
and its respective eigenvector is given by 
\begin{equation}
|FR_{2}\rangle=\prod_{i=1}^{N}\left|2,0\right\rangle _{i}|-\rangle_{i}.\label{Trt-FR2}
\end{equation}
The definition of two-fold degenerate eigenstate $\left|2,0\right\rangle _{i}$
is reported in Table~\ref{tab1}, which implies that the second frustrated
phase $FR_{2}$ also has residual entropy $\mathcal{S}=\ln(2)$. The
Ising spin magnetization is $m_{I}=-\frac{1}{2}$, the Heisenberg
spin magnetization is $m_{H}=2$, and the total magnetization results
in $m_{t}=\tfrac{3}{2}$.

The ground-state energy of the third frustrated phase $FR_{3}$ follows
from the relation 
\begin{alignat}{1}
E_{_{FR_{3}}}=-2J_{0}+ & J-J_{z}-2h_{z}-\tfrac{1}{2}h,
\end{alignat}
whereas its respective eigenvector reads 
\begin{equation}
|FR_{3}\rangle=\prod_{i=1}^{N}\left|2,0\right\rangle _{i}|+\rangle_{i}.\label{Trt-FR3}
\end{equation}
The two-fold degenerate eigenvector $\left|2,0\right\rangle _{i}$
is defined in Table~\ref{tab1} and hence, the third frustrated phase
$FR_{2}$ is macroscopically degenerate with the residual entropy
$\mathcal{S}=\ln(2)$ per unit cell. The corresponding Ising spin
magnetization achieves the value $m_{I}=\frac{1}{2}$, the Heisenberg
spin magnetization equals to $m_{H}=2$, and the total magnetization
is given by $m_{t}=\tfrac{5}{2}$.

Usually, plots can be drawn in units of some parameters
like $J$, and then the temperature can be measured in units $J$.
However, here for convenience, we set the parameters to be $J=-10$
and $J_{0}=-10$, just for scale the temperature by a factor $10$.
From now on, we will consider this set of parameters to study the
pseudo-critical temperature throughout the article.

All dashed lines in Fig.~\ref{fig:Zero-temperature}(a) represent
usual ground-state phase boundaries between two phases. The residual
entropy per unit cell at the phase boundary between $FR_{1}$ and
$FR_{2}$ becomes $\mathcal{S}=\ln(4)$. Similarly, the residual entropy
at the interface between $FR_{2}$ and $FI_{3}$ equals to $\mathcal{S}=\ln(3)$,
while the residual entropy at the phase boundary between $FI_{3}$
and $SA$ equals to $\mathcal{S}=\ln(2)$. Analogously, the residual
entropy attains the value $\mathcal{S}=\ln(3)$ at phase boundaries
between $SA-FR_{3}$ and $FR_{3}-FI_{2}$. Finally, the residual entropy
becomes $\mathcal{S}=\ln(2)$ at the interface between $FI_{2}$ and
$FI_{3}$. In all aforementioned cases the residual entropy per unit
cell is always higher than the entropy of both individual phases,
which coexist together at a relevant ground-state boundary. By contrast,
solid lines represent all unusual phase boundaries between two phases.
The residual entropy per unit cell $\mathcal{S}=\ln(2)$ can be found
at interfaces between the phases $FR_{1}$-$FI_{1}$, $FR_{2}$-$FI_{1}$,
$FR_{2}$-$FI_{2}$, and $FR_{3}$-$FI_{3}$, whereas the residual
entropy per unit cell vanishes $\mathcal{S}=0$ at the interface between
two non-degenerate ferrimagnetic phases $FI_{2}$ and $FI_{3}$.

\section{Thermodynamics}

\label{Sec3}

The mixed spin-(1/2,1) Ising-Heisenberg double-tetrahedral chain can
be mapped onto the effective spin-1/2 Ising chain given by the Hamiltonian
\begin{equation}
H=-\sum_{i=1}^{N}\left[K_{0}+Ks_{i}s_{i+1}+\tfrac{1}{2}B(s_{i}+s_{i+1})\right],
\end{equation}
where $K_{0}$, $K$, and $B$ are effective temperature-dependent
parameters. Bearing this in mind, thermodynamics of the effective
spin-1/2 Ising chain can be expressed in terms of the transfer matrix
$\mathbf{V}=\left[\begin{array}{cc}
w_{1} & w_{0}\\
w_{0} & w_{-1}
\end{array}\right]$ according to the procedure previously discussed in Ref.~\citep{pseudo}.
Each element of the transfer matrix (Boltzmann factor) $w_{n}$ with
$n=\{-1,0,1\}$, which will be further referred to as the sector,
can be defined as 
\begin{equation}
w_{n}=\sum_{k=0}^{18}g_{n,k}\,{\rm e}^{-\beta\varepsilon_{n,k}},\label{eq:G-w}
\end{equation}
where $\beta=1/(k_{B}T)$, $k_{B}$ is Boltzmann's constant, $T$
is the absolute temperature and the eigenvalues $\varepsilon_{n,k}$
are given by Eq.~\eqref{eq:E-spct}.

To be more specific, the Boltzmann factors are explicitly given by
\begin{alignat}{1}
w_{n}= & u^{n}\left\{ q_{3,n}\,z^{6}+\left(x^{4}+\frac{2}{x^{2}}\right)z^{2}\,q_{2,n}+\frac{\left(2t+x^{-4}\right)}{z^{2}}\right.\nonumber \\
 & \left.+\frac{1}{z}\left[\left(\frac{2\,y_{1}}{x}+x^{2}\,y_{2}\right)q_{1,n}+x^{2}\,y_{3}\right]\right\} ,
\end{alignat}
where $x={\rm e}^{\beta J/2}$, $z={\rm e}^{\beta J_{z}/2}$, $u={\rm e}^{\beta h/2}$,
$t=2\cosh\left(\beta J\right)$, while the coefficients $y_{r}$ and
$q_{r,n}$ with $r=\{1,2,3\}$ are defined as follows 
\begin{alignat}{1}
y_{r}= & 2\cosh\left[\beta J\csc\left(2\phi_{r}\right)\right],\\
q_{r,n}= & 2\cosh\left[r\beta\left(nJ_{0}+h_{z}\right)\right].
\end{alignat}

The transfer-matrix eigenvalues are determined by the following equation
\begin{equation}
\lambda_{\pm}=\tfrac{1}{2}\Bigl(w_{1}+w_{-1}\pm\sqrt{(w_{1}-w_{-1})^{2}+4w_{0}^{2}}\Bigr).\label{eq:L12}
\end{equation}
Considering the effective spin-1/2 Ising chain under a periodic boundary
condition gives the partition function $\mathcal{Z}_{N}=\lambda_{+}^{N}+\lambda_{-}^{N}$.
Consequently, the free energy can be obtained in the thermodynamic
limit ($N\rightarrow\infty$) according to the formula 
\begin{equation}
f=-\tfrac{1}{\beta}\ln\left[\tfrac{1}{2}\Bigl(w_{1}+w_{-1}+\sqrt{(w_{1}-w_{-1})^{2}+4w_{0}^{2}}\Bigr)\right].\label{eq:free-energ}
\end{equation}
Substituting Boltzmann's factors $w_{n}$ into Eq.~\eqref{eq:free-energ},
we can exactly calculate the free energy of the mixed spin-(1/2,1)
Ising-Heisenberg double-tetrahedral chain at finite temperature.

It has been recently demonstrated \citep{pseudo} that some 1D lattice-statistical
models satisfy the following condition $|w_{1}-w_{-1}|\gg w_{0}$
at low enough temperatures. Under this condition, the free energy
of the mixed spin-(1/2,1) Ising-Heisenberg double-tetrahedral chain
reduces to 
\begin{equation}
f=-T\ln\left\{ \max\left[w_{1}(T),w_{-1}(T)\right]\right\} .\label{eq:f-max}
\end{equation}
The final formula for the free energy per unit cell \eqref{eq:free-energ}
takes the following simple form at a phase boundary between the individual
phases with the same energy $\varepsilon_{c}$ 
\begin{equation}
f=\varepsilon_{c}-T\ln\left[\max\left(g_{1,0},g_{-1,0}\right)\right].\label{eq:free-dsc}
\end{equation}
Consequently, the residual entropy per unit cell at a relevant phase
boundary reads 
\begin{equation}
\mathcal{S}_{c}=\ln\left[\max\left(g_{1,0},g_{-1,0}\right)\right].\label{eq:S_c}
\end{equation}
Knowing this quantity is sufficient for prediction of a pseudo-transition
at finite temperatures \citep{ph-bd}.

In Fig.~\ref{fig:Zero-temperature}(b) we illustrate the density
plot of the entropy as a function of $J_{z}$ and $h$ for the fixed
temperature $T=0.4$ by using the same scale as in the ground-state
phase diagram shown in Fig.~\ref{fig:Zero-temperature}(a). It is
quite evident that the entropy follows the vestige of zero-temperature
phase diagram at finite temperatures. The notation for the ground
state is changed at finite temperatures by adding a prefix \textquotedbl$q$\textquotedbl{}
to the name of respective ground states, which will denote the respective
quasi-phase \citep{Timonin} because of a lack of true spontaneous
long-range order at finite temperatures. It could be expected that
thermal excitations basically influence the phase boundaries. It has
been argued previously that all dashed curves displayed in Fig.~\ref{fig:Zero-temperature}(a)
describe standard interfaces, which are manifested through an increase
of the entropy exceeding the entropy value of both coexisting phases.
Contrary to this, the phase boundaries depicted by solid lines in
Fig.~\ref{fig:Zero-temperature}(a) behave quite differently, since
they show at the respective interface a sharp rise of the entropy
to a greater entropy of one of two coexisting phases.

\begin{figure}[h]
\centering{}\includegraphics[scale=0.43]{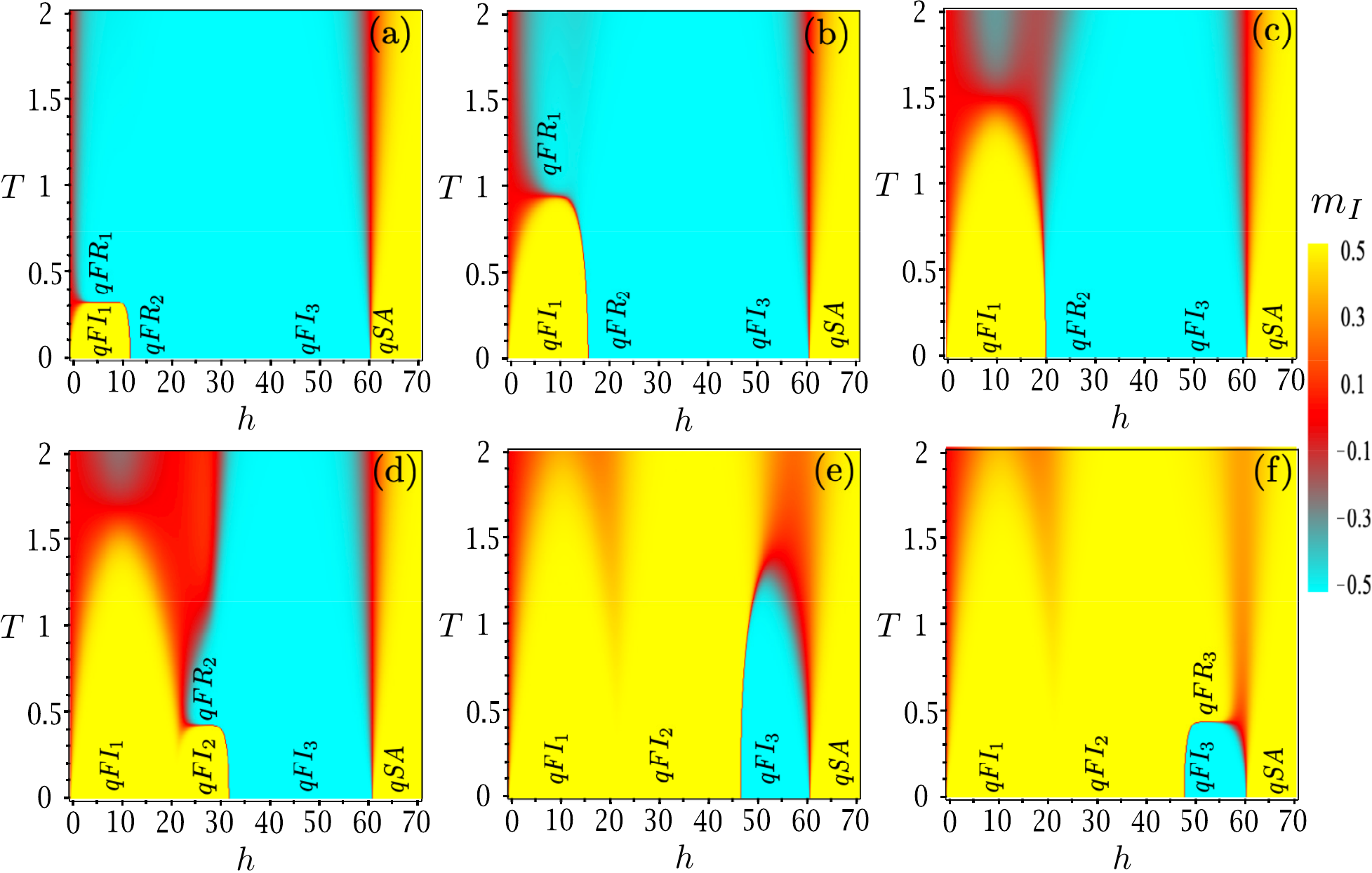} \caption{\label{fig:phs-dgm-T} Density plot of Ising spin magnetization in
the $T-h$ plane for the fixed values of the coupling constants $J=-10$,
$J_{0}=-10$, and several values of $J_{z}$: (a) $J_{z}=-11$; (b)
$J_{z}=-13$; (c) $J_{z}=-15$; (d)$J_{z}=-15.65$; (e) $J_{z}=-19$;
(f) $J_{z}=-19.85$.}
\end{figure}

The density plot of Ising spin magnetization is depicted in Fig.~\ref{fig:phs-dgm-T}
in the $T-J_{z}$ plane for the following set of parameters $J=-10$
and $J_{0}=-10$. In this figure, yellow region corresponds to spin
'up' ($m_{I}=1/2$), cyan region corresponds to spin 'down' ($m_{I}=-1/2$),
and red region corresponds to null Ising magnetization ($m_{I}=0$).
Surely the temperature in units of $T/|J|$ would be divided by a
factor 10 in Fig.~\ref{fig:phs-dgm-T} and the following figures.

\begin{figure}[h]
\centering{} \includegraphics[scale=0.43]{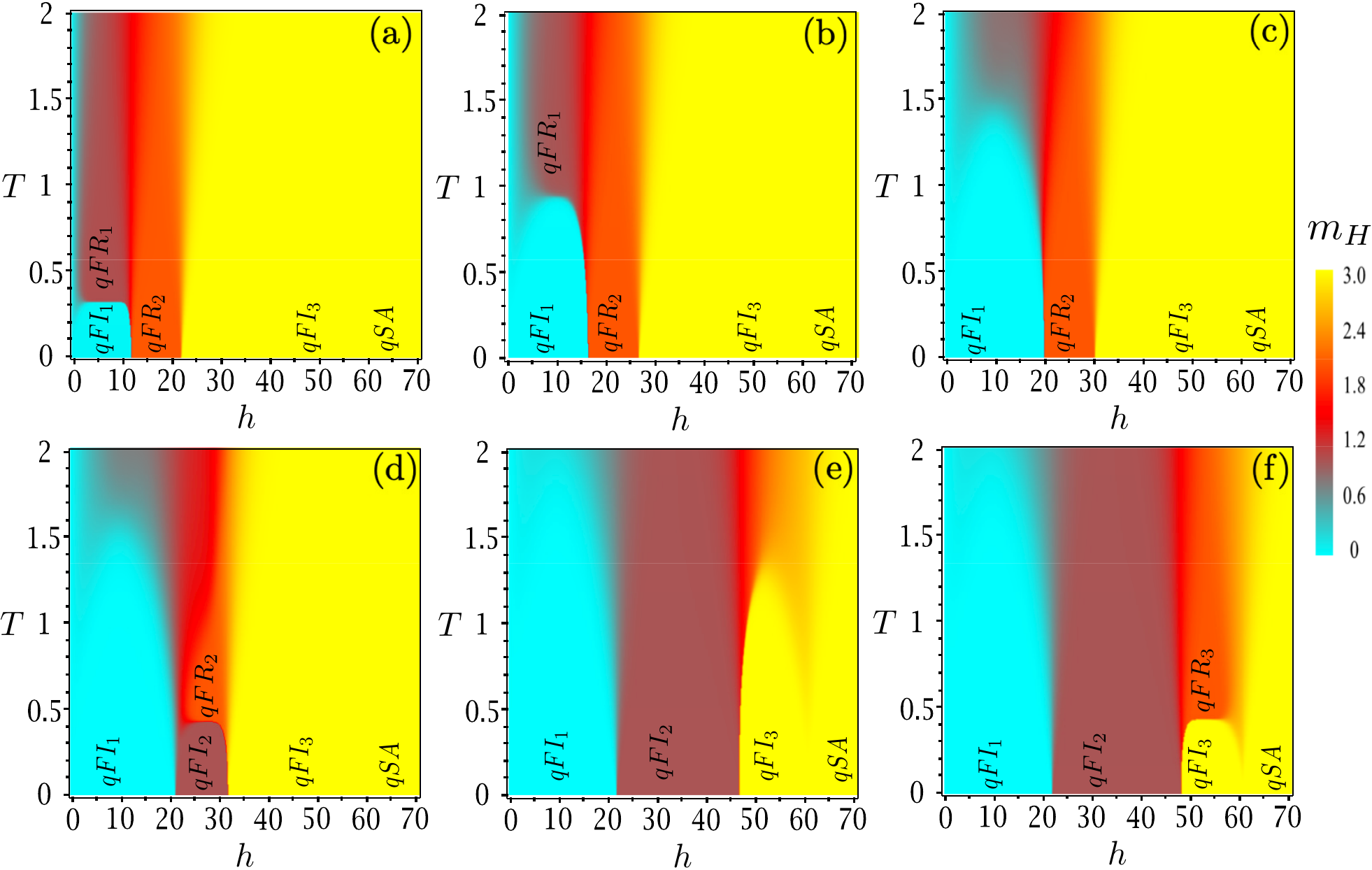} \caption{\label{fig:phs-dgm-Hsg} Density plot of Heisenberg spin magnetization
in the $T-h$ plane for the fixed values of the coupling constants
$J=-10$, $J_{0}=-10$, and several values of $J_{z}$: (a) $J_{z}=-11$;
(b) $J_{z}=-13$; (c) $J_{z}=-15$; (d)$J_{z}=-15.65$; (e) $J_{z}=-19$;
(f) $J_{z}=-19.85$.}
\end{figure}

The density plot of the Heisenberg spin magnetization is depicted
in Fig.~\ref{fig:phs-dgm-Hsg} in the $T-J_{z}$ plane for the same
set of parameters $J=-10$ and $J_{0}=-10$. The color code for the
density plot is as follows: yellow region corresponds to the saturated
Heisenberg magnetization $m_{H}=3$, cyan region corresponds to the
null Heisenberg magnetization $m_{H}=0$, orange region corresponds
to the moderate Heisenberg magnetization $m_{H}=2$, and dark red
region corresponds to the moderate Heisenberg magnetization $m_{H}=1$.
It can be seen from Figs.~\ref{fig:phs-dgm-T} and \ref{fig:phs-dgm-Hsg}
that the pseudo-transitions between the quasi-phases is accompanied
with abrupt change in the magnetization of the Ising spins and/or
the magnetization of the Heisenberg spins. The density plots shown
in Fig.~\ref{fig:phs-dgm-Hsg}(a)-(f) imply a full alignment of the
Heisenberg spins just within the quasi-phases $qFI_{3}$ and $qSA$.

\begin{figure}
\includegraphics[scale=0.45]{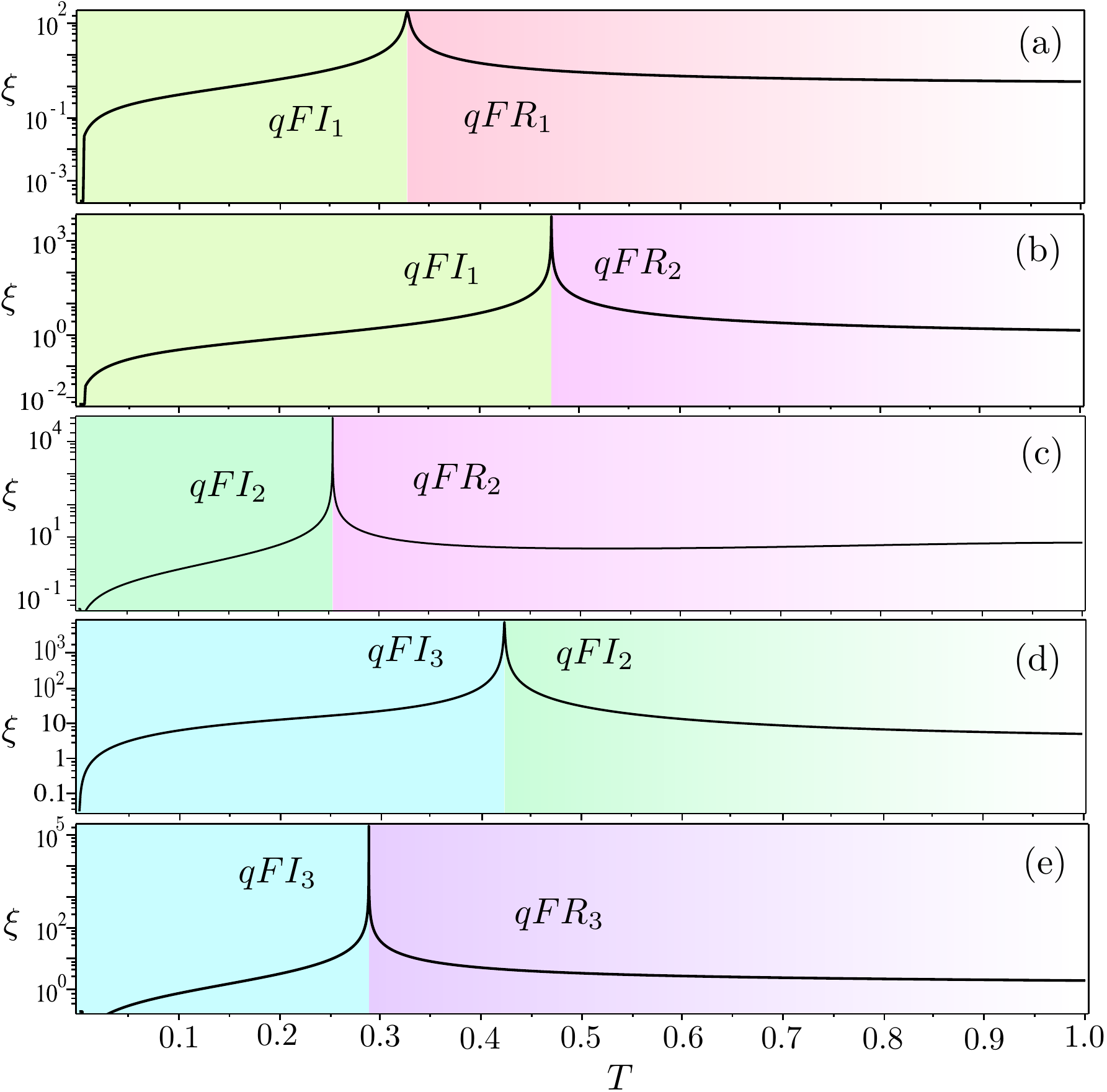} \caption{\label{fig:Correlation-length} Correlation length against temperature
for the fixed parameters $J=-10$, $J_{0}=-10$ and several values
of $J_{z}$ and $h_{z}=h$: (a) $h=4$, $J_{z}=-11$; (b) $h=11$;
$J_{z}=-11.5$; (c) $h=26$, $J_{z}=-15.6$; (d) $h=36.76$, $J_{z}=-17$;
(e) $h=52$; $J_{z}=-19.9$.}
\end{figure}

Now, let us analyze the correlation length, which can be calculated
according to the following simple relation 
\begin{equation}
\xi=\left[\ln\left(\frac{\lambda_{+}}{\lambda_{-}}\right)\right]^{-1}.
\end{equation}
The correlation length is depicted in Fig.~\ref{fig:Correlation-length}
as a function of temperature for the fixed parameters $J=-10$, $J_{0}=-10$,
and $h_{z}=h$. It is advisable to follow the zero-temperature phase
diagram to interpret the relevant dependences of the correlation length.
In Fig.~\ref{fig:Correlation-length}(a) we illustrate the correlation
length for $h=4$ and $J_{z}=-11$, whereas the shark peak delimits
the quasi-phases $qFI_{1}$ and $qFR_{1}$ in agreement with the ground-state
phase phase diagram shown in Fig.~\ref{fig:Zero-temperature}(a).
Although the correlation length seems to diverge at a pseudo-critical
temperature, it is in fact just a sharp finite peak. In Fig.~\ref{fig:Correlation-length}(b)
one observes a similar curve for $h=11$ and $J_{z}=-11.5$, but now
the peak indicates a pseudo-transition between the quasi-phases $qFI_{1}$
and $qFR_{2}$. Fig.~\ref{fig:Correlation-length}(c) depicts the
correlation length for $h=30$ and $J_{z}=-15.65$, whereas the sharp
peak determines a pseudo-transition between the quasi-phases $qFI_{2}$
and $qFR_{2}$. Similarly, the correlation length plotted in Fig.~\ref{fig:Correlation-length}(d)-(e)
demonstrates that a pseudo-transition between the quasi-phases $qFI_{3}$-$qFI_{2}$
and $qFI_{3}$-$qFR_{3}$ are accompanied with a sharp robust peak
of the correlation length. It is worthy to mention that the quasi-phases
melt smoothly upon increasing temperature when the temperature is
higher than the pseudo-critical temperature.

It is quite clear from Eq.~\eqref{eq:f-max} that the pseudo-critical
temperature $T_{p}$ can be alternatively obtained by solving the
equation 
\begin{equation}
w_{1}(T_{p})=w_{-1}(T_{p}).\label{eq:w-Tp}
\end{equation}
\begin{widetext} 
\begin{figure}[h]
\includegraphics[scale=0.9]{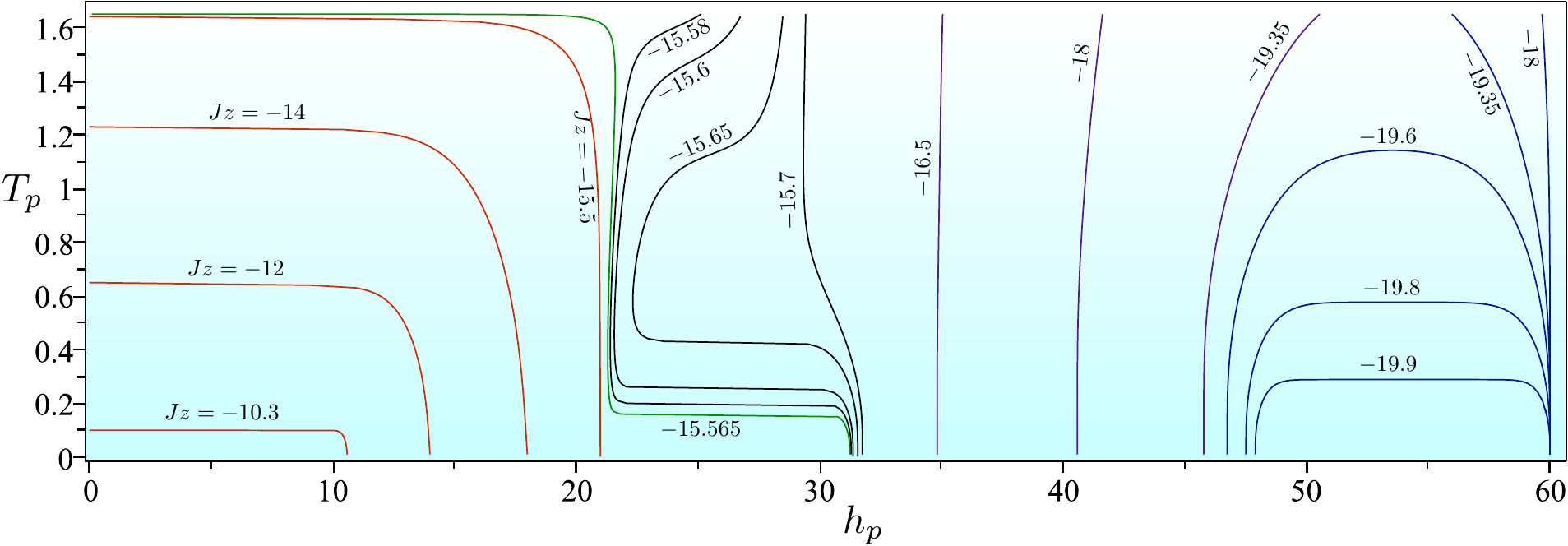} \caption{\label{fig:Psd-crit-temp-h} Pseudo-critical temperature as a function
of the magnetic field for the fixed values of interaction parameters
$J=-10$, $J_{0}=-10$, $h_{z}=h$ and several values of $J_{z}$.}
\end{figure}

\end{widetext} The numerical solution of Eq.~(\ref{eq:w-Tp}) allows
us to plot the pseudo-critical temperature $T_{p}$ against the magnetic
field $h_{p}$ for several values of $J_{z}$ (see Fig.~\ref{fig:Psd-crit-temp-h}).
For sufficiently low magnetic fields $0<h_{p}<10$ the pseudo-critical
temperature delimits the quasi-phases $qFI_{1}$ (below the curve)
and $qFR_{1}$ (above the curve), whereas for the moderate fields
$10\lesssim h_{p}\lesssim21$ the pseudo-transition line delimits
the quasi-phases $qFI_{1}$ (left from the curve) and $qFR_{2}$ (right
from the curve). Furthermore, the investigated model undergoes a pseudo-transition
between the quasi-phases $qFI_{2}$ and $qFR_{2}$ for $T_{p}\lesssim0.6$
and $21\lesssim h_{p}\lesssim31$, while the pseudo-transition between
the quasi-phases $qFI_{2}$ (left side of the curve) and $qFI_{3}$
(right side of the curve) takes place for $31\lesssim h_{p}\lesssim51$.
Finally, the pseudo-transition line delimits the quasi-phases $qFI_{3}$
(below the curve) and $qFR_{3}$ (above the curve) for high enough
magnetic fields $51\lesssim h_{p}<60$. Although the condition \eqref{eq:w-Tp}
may still give relatively high values of the pseudo-critical temperature
(e.g., $T\gtrsim1$), it turns out that the pseudo-critical line melts
smoothly for sufficiently high temperatures $T\sim1$ (in some particular
cases even at lower temperatures). In general, there is no way to
identify the maximum value of the pseudo-critical temperature. Besides,
the pseudo-critical temperature also melts for $h_{p}\rightarrow0$
and $h_{p}\rightarrow60$ as evidenced by Figs. \ref{fig:phs-dgm-T}
and \ref{fig:phs-dgm-Hsg}.

\begin{figure}[h]
\includegraphics[scale=0.45]{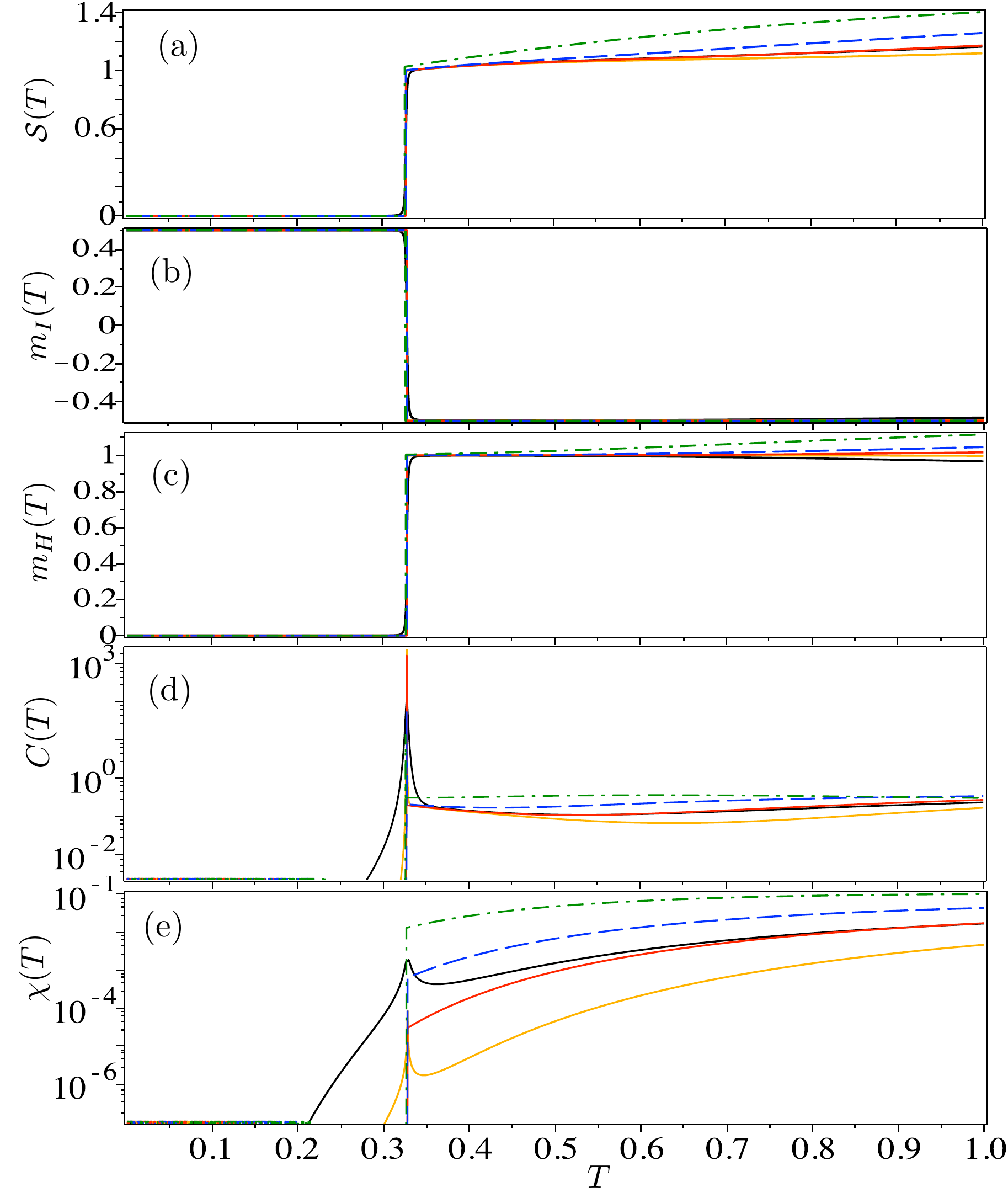} \caption{\label{fig:Phs-qnty-T1} Temperature dependences of some thermodynamic
quantities by considering the fixed parameters $J=-10$, $J_{0}=-10$,
$J_{z}=-11$, and several values of the magnetic field $h=\{4,6,8,9,10\}$
(black solid, orange solid, red solid, blue dashed, and green dot
dashed): (a) entropy ${\cal S}$; (b) Ising spin magnetization; (c)
Heisenberg spin magnetization; (d) specific heat (semi-logarithmic
plot); (e) magnetic susceptibility (semi-logarithmic plot).}
\end{figure}

Temperature variations of some thermodynamic quantities are plotted
in Fig.~\ref{fig:Phs-qnty-T1} close to a pseudo-transition between
the quasi-phases $qFR_{1}$ and $qFI_{1}$ for the fixed values of
the interaction parameters $J=-10$, $J_{0}=-10$, $J_{z}=-11$, and
several values of the magnetic field $h=\{4,6,8,9,10\}$ outlined
by \{black solid, orange solid, red solid, blue dashed, and green
dot dashed\} curves, respectively. A strong thermally-induced change
of the entropy ${\cal S}(T)$ is observable in Fig.~\ref{fig:Phs-qnty-T1}(a)
around the pseudo-critical temperature $T_{p}\approx0.3275$. It is
worthy to mention that the pseudo-critical temperature remains almost
constant for $0<h<10$. It is quite evident from Fig.~\ref{fig:Phs-qnty-T1}(b),
moreover, that the Ising spins are mostly aligned parallel to the
magnetic field ($m_{I}=0.5$) below the pseudo-critical temperature
$T<T_{p}$ and antiparallel ($m_{I}=-0.5$) above it $T>T_{p}$. Contrary
to this, the Heisenberg spins almost do not contribute to the total
magnetization ($m_{H}=0$) below the pseudo-critical temperature $T<T_{p}$,
while they provide a significant contribution $(m_{H}=1)$ above it
$T>T_{p}$. Last but not least, the specific heat and magnetic susceptibility
displayed in Fig.~\ref{fig:Phs-qnty-T1}(d)-(e) in a semi-logarithmic
scale serve in evidence of a pseudo-transition through a strong narrow
peak observable at the pseudo-critical temperature.

\begin{figure}[h]
\includegraphics[scale=0.45]{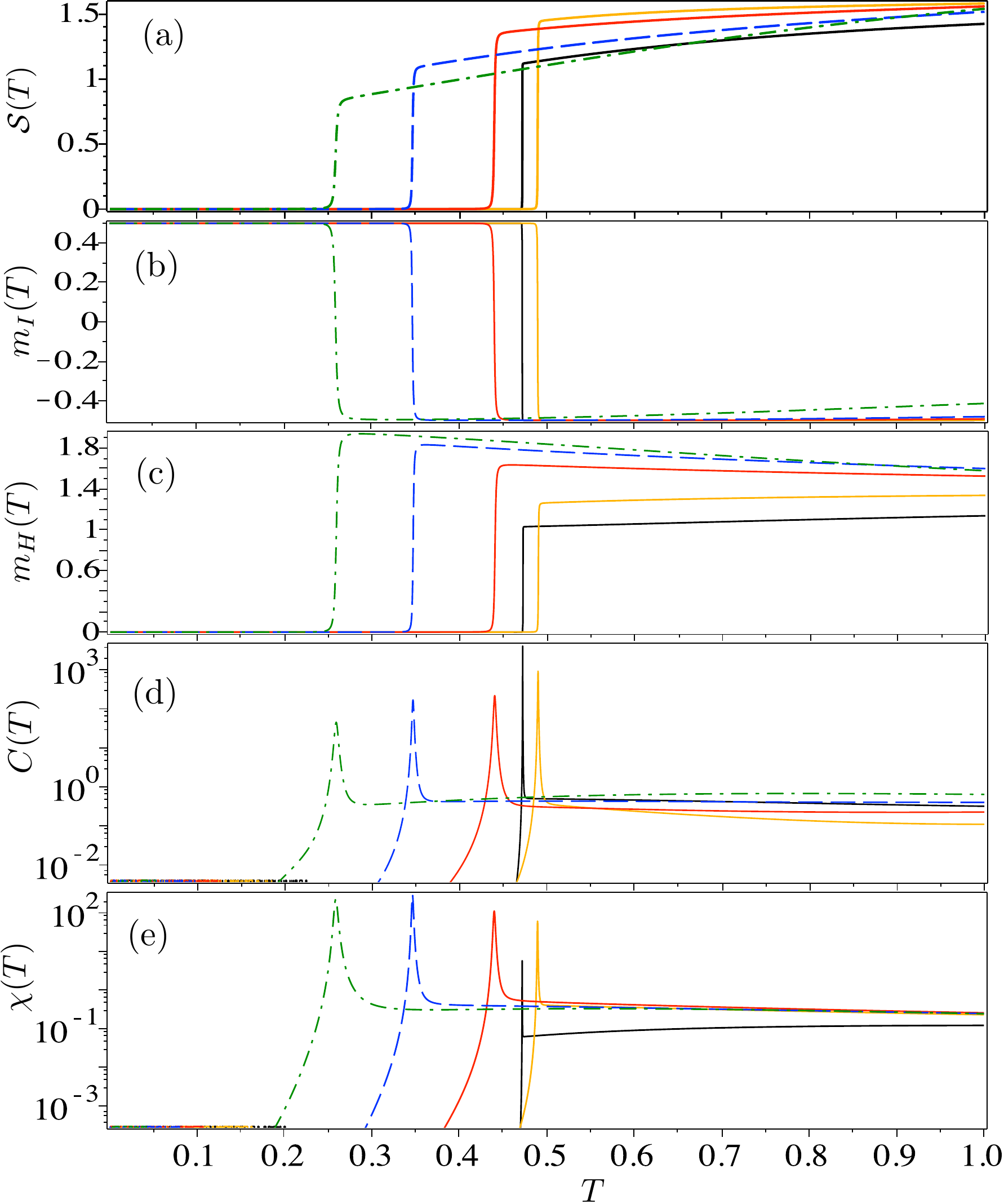} \caption{\label{fig:Phs-qnty-T2} Temperature dependences of some thermodynamic
quantities by considering the fixed parameters $J=-10$, $J_{0}=-10$,
and $(h,J_{z})=$\{$(11,-11.5)$, $(13,-12)$, $(15.5,-13)$, $(16.9,13.6)$,
$(18.81,-14.5)$\} (black solid, orange solid, red solid, blue dashed,
and green dot dashed): (a) entropy ${\cal S}$; (b) Ising spin magnetization;
(c) Heisenberg spin magnetization; (d) specific heat (semi-logarithmic
plot); (e) magnetic susceptibility (semi-logarithmic plot).}
\end{figure}

Temperature dependences of selected thermodynamic quantities are depicted
in Fig.~\ref{fig:Phs-qnty-T2} by assuming the fixed values of the
interaction parameters $J=-10$, $J_{0}=-10$, and $(h,J_{z})=$\{$(11,-11.5)$,
$(13,-12)$, $(15.5,-13)$, $(16.9,13.6)$, $(18.81,-14.5)$\} outlined
by \{black solid, orange solid, red solid, blue dashed, and green
dot dashed\} curves, respectively. The present choice of the interaction
parameters is consistent with the pseudo-transition between the quasi-phases
$qFI_{1}$ and $qFR_{2}$, which varies with the interaction parameter
$J_{z}$ and magnetic field $h$. It is obvious from Fig.~\ref{fig:Phs-qnty-T2}(a)
that the entropy ${\cal S}(T)$ exhibits a steep increase close to
a pseudo-critical temperature $T_{p}$, while the magnetization of
Ising spins shown in Fig.~\ref{fig:Phs-qnty-T2}(b) is pointing upward
($m_{I}=0.5$) for $T<T_{p}$ and downward ($m_{I}=-0.5$) for $T>T_{p}$.
Similarly, the magnetization of Heisenberg spins illustrated in Fig.
\ref{fig:Phs-qnty-T2}(c) is zero ($m_{H}=0$) for $T<T_{p}$, while
there is a sudden change at $T=T_{p}$ above which it strongly depends
on the magnetic field $h$ and the coupling constant $J_{z}$. Finally,
sharp narrow peaks can be repeatedly detected at a pseudo-critical
temperature in the respective temperature dependences of the specific
heat {[}Fig.~\ref{fig:Phs-qnty-T2}(d){]} and the magnetic susceptibility
{[}Fig.~\ref{fig:Phs-qnty-T2}(e){]}.

\begin{figure}[h]
\includegraphics[scale=0.45]{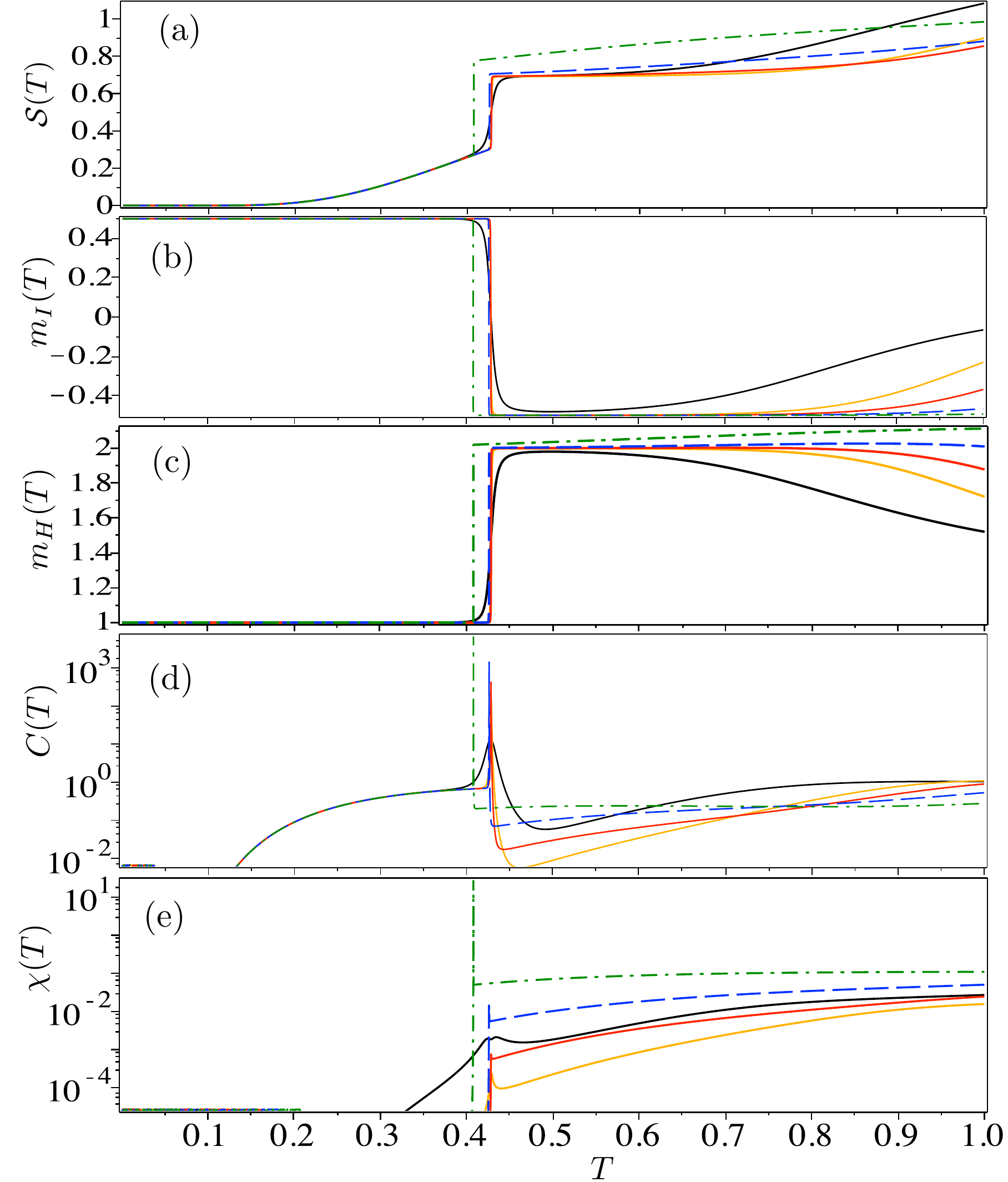} \caption{\label{fig:Phs-qnty-T3} Temperature dependences of some thermodynamic
quantities by considering the fixed parameters $J=-10$, $J_{0}=-10$,
$J_{z}=-15.65$, and several values of the magnetic field $h=\{25,27,28,29,30\}$
(black solid, orange solid, red solid, blue dashed, and green dot
dashed): (a) entropy ${\cal S}$; (b) Ising spin magnetization; (c)
Heisenberg spin magnetization; (d) specific heat (semi-logarithmic
plot); (e) magnetic susceptibility (semi-logarithmic plot).}
\end{figure}

A pseudo-transition between the quasi-phases $qFI_{2}$ and $qFR_{2}$
is illustrated in Fig.~\ref{fig:Phs-qnty-T3} by considering the
fixed parameters $J=-10$, $J_{0}=-10$, $J_{z}=-15.65$, and several
values of the magnetic field of $h=\{25,27,28,29,30\}$ outlined by
\{black solid, orange solid, red solid, blue dashed, and green dot
dashed\} curves, respectively. Fig.~\ref{fig:Phs-qnty-T3}(a) shows
the entropy ${\cal S}(T)$ as a function of temperature: for $T<T_{p}$
the entropy increases significantly but is virtually independent of
$h$ (for $22\apprle h\apprle30$), then a sudden rise occurs at $T=T_{p}$
followed by a successive smooth increase for $T>T_{p}$. The Ising
magnetization depicted in Fig.~\ref{fig:Phs-qnty-T3}(b) is nearly
constant $m_{I}=0.5$ for $T<T_{p}$, but it becomes almost $-0.5$
for $T\gtrsim T_{p}$ before showing a continuous rise approaching
null upon further increase of temperature. Analogously, the Heisenberg
spin magnetization illustrated in Fig.~\ref{fig:Phs-qnty-T3}(c)
tends to zero $m_{H}\rightarrow1$ for $T<T_{p}$, while it approaches
to $m_{H}\rightarrow2$ for $T\gtrsim T_{p}$. The specific heat and
magnetic susceptibility plotted in Fig.~\ref{fig:Phs-qnty-T3}(d)-(e)
in a semi-logarithmic scale display vigorous narrow peaks verifying
a pseudo-transition between the quasi-phases $qFI_{2}$ and $qFR_{2}$.

\begin{figure}[h]
\includegraphics[scale=0.45]{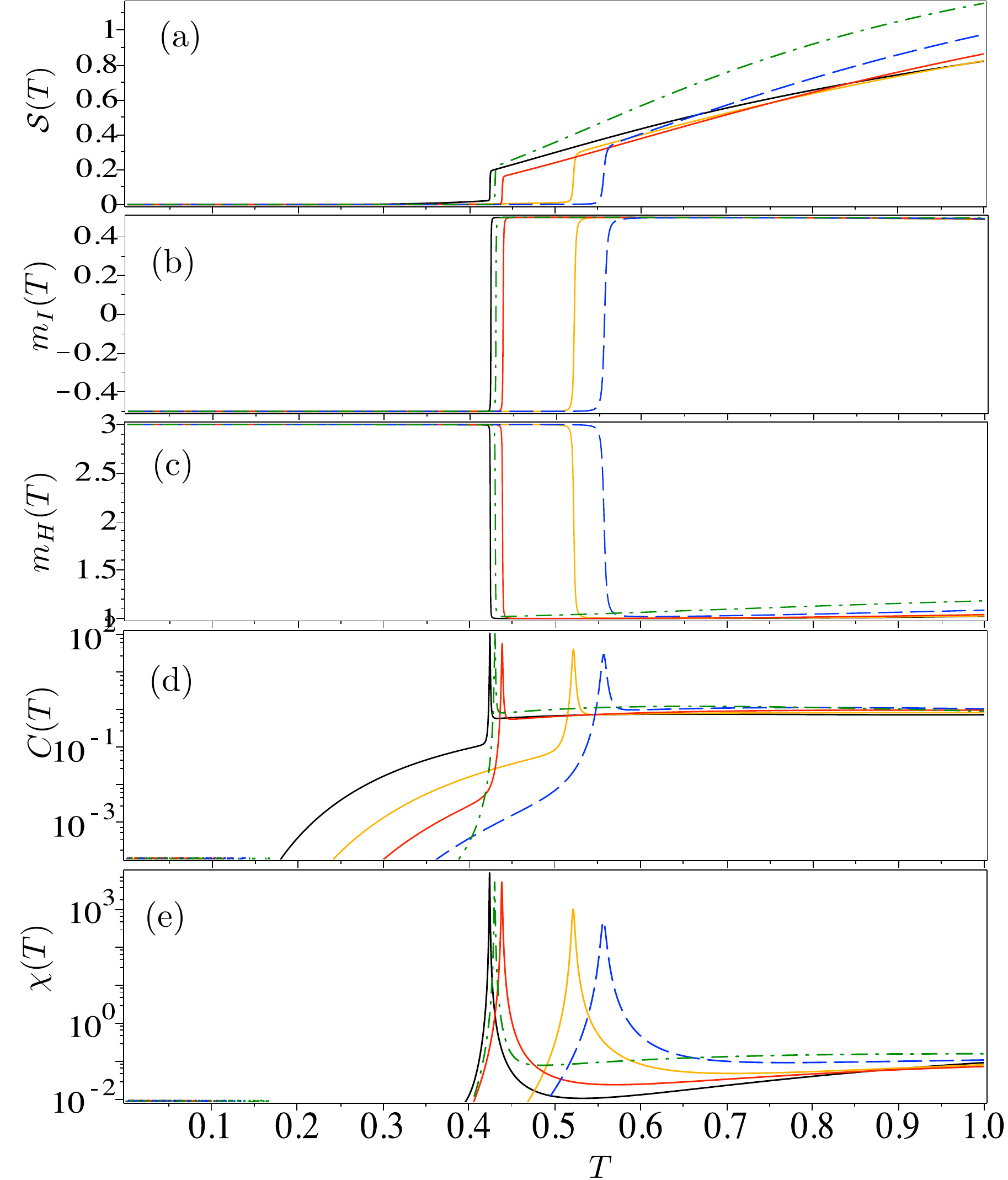} \caption{\label{fig:Phs-qnty-T4} Temperature dependences of some thermodynamic
quantities by considering the fixed parameters $J=-10$, $J_{0}=-10$,
and $(h,J_{z})=$\{$(36.76,-17)$, $(38.7,-17.5)$, $(40.6,-18)$,
$(42.55,18.5)$, $(44.45,-19)$\} (black solid, orange solid, red
solid, blue dashed, and green dot dashed)): (a) entropy ${\cal S}$;
(b) Ising spin magnetization; (c) Heisenberg spin magnetization; (d)
specific heat (semi-logarithmic plot); (e) magnetic susceptibility
(semi-logarithmic plot).}
\end{figure}

Next, the pseudo-transition at the interface between the quasi-phases
$qFI_{2}$ and $qFI_{3}$ is illustrated in Fig.~\ref{fig:Phs-qnty-T4}
by considering set of the parameters $J=-10$, $J_{0}=-10$, and $(h,J_{z})=$\{$(36.76,-17)$,
$(38.7,-17.5)$, $(40.6,-18)$, $(42.55,18.5)$, $(44.45,-19)$\}
drawn by \{black solid, orange solid, red solid, blue dashed, and
green dot dashed\} curves, respectively. It should be stressed that
both coexisting quasi-phases $qFI_{2}$ and $qFI_{3}$ are non-frustrated
and consequently, the residual entropy per unit cell should also become
null according to Eq.~\eqref{eq:S_c}. The entropy ${\cal S}(T)$
as a function of temperature shown in Fig.~\ref{fig:Phs-qnty-T4}(a)
is for $T<T_{p}$ nearly zero, then it shows a small but sudden rise
at $T=T_{p}$, which is followed by a roughly linear increase for
$T>T_{p}$. The magnetization of Ising spins {[}Fig.~\ref{fig:Phs-qnty-T4}(b){]}
displays an opposite behavior to the previous one: the Ising spins
are aligned in opposite to the magnetic field ($m_{I}=-0.5$) for
$T<T_{p}$ and they are aligned in the magnetic-field direction ($m_{I}=0.5$)
for $T>T_{p}$. Similarly, the magnetization of Heisenberg spins {[}Fig.~\ref{fig:Phs-qnty-T4}(c){]}
is close to its maximal value $m_{H}=3$ for $T<T_{p}$ and it suddenly
drops to $m_{H}=1$ for $T>T_{p}$. Finally, one observes a typical
narrow peak in thermal variations of the specific heat and magnetic
susceptibility displayed in Fig.~\ref{fig:Phs-qnty-T4}(d)-(e).

\begin{figure}[h]
\includegraphics[scale=0.45]{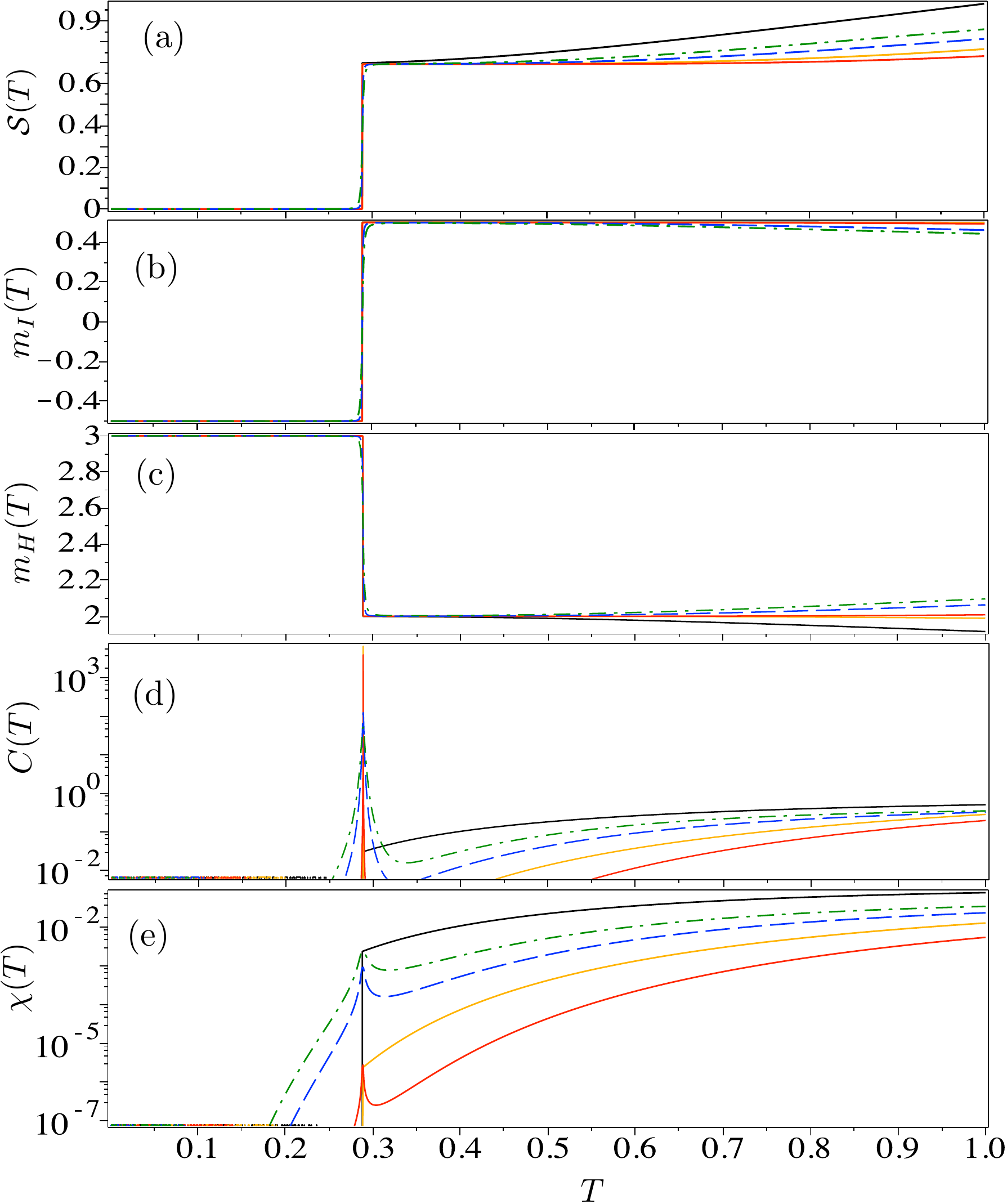} \caption{\label{fig:Phs-qnty-T5} Temperature dependences of some thermodynamic
quantities by considering the fixed parameters $J=-10$, $J_{0}=-10$,
$J_{z}=-19.9$, and several values of the magnetic field $h=\{50,52,54,56,56.5\}$
(black solid, orange solid, red solid, blue dashed, and green dot
dashed): (a) entropy ${\cal S}$; (b) Ising spin magnetization; (c)
Heisenberg spin magnetization; (d) specific heat (semi-logarithmic
plot); (e) magnetic susceptibility (semi-logarithmic plot).}
\end{figure}

Last but not least, let us discuss a pseudo-transition between the
quasi-phases $qFR_{3}$ and $qFI_{3}$ exemplified in Fig.~\ref{fig:Phs-qnty-T5}
for the fixed values of the interaction parameters $J=-10$, $J_{0}=-10$,
$J_{z}=-19.9$, and several magnetic fields $h=\{50,52,54,56,56.5\}$
sketched by \{black solid, orange solid, red solid, blue dashed, and
green dot dashed\} curves, respectively. It is noteworthy that thermal
variation of the entropy ${\cal S}(T)$ displayed in Fig.~\ref{fig:Phs-qnty-T5}(a)
is quite reminiscent of the entropy dependence illustrated in Fig.~\ref{fig:Phs-qnty-T1}(a).
In addition, the temperature dependences of the magnetization of the
Ising and Heisenberg spins shown in Fig.~\ref{fig:Phs-qnty-T5}(b)
and (c) are quite similar to the previous cases shown in Fig.~\ref{fig:Phs-qnty-T4}(b)
and (c), respectively. Although the specific heat shows a strong narrow
peak at the pseudo-critical temperature, it often becomes negligible
further away from the pseudo-critical temperature {[}see Fig.~\ref{fig:Phs-qnty-T5}(d){]}.
The similar situation can be also found in the temperature dependences
of the magnetic susceptibility shown in Fig.~\ref{fig:Phs-qnty-T5}(e).

\section{Conclusions}

\label{Sec4}

The pseudo-transitions of the mixed spin-(1/2,1) Ising-Heisenberg
double-tetrahedral chain are examined in detail at non-zero temperature
and magnetic field. The ground-state phase diagram of the investigated
spin chain totally involves seven phases, three of which can be classified
as the non-degenerate ferrimagnetic phases, three as the macroscopically
degenerate frustrated phases, and one as the saturated paramagnetic
phase. Interestingly, five different ground-state boundaries of the
mixed spin-(1/2,1) Ising-Heisenberg double-tetrahedral chain represent
peculiar interfaces, at which the residual entropy per unit cell is
simply given by the larger entropy of one of two coexisting phases.
This condition seems to be sufficient criterion whether or not the
pseudo-transition does emerge in a close vicinity of the ground-state
phase boundary. In fact, the residual entropy per unit cell at the
usual ground-state phase boundaries is strictly larger than the residual
entropy of both coexisting phases. Although thermal fluctuations usually
destroy in 1D lattice-statistical models with short-range interactions
all fingerprints of the ground-state phase boundaries, the aforementioned
five interfaces are quite robust with respect to thermal fluctuations.
In consequence of that, the mixed spin-(1/2,1) Ising-Heisenberg double-tetrahedral
chain may exhibit in a vicinity of five aforedescribed ground-state
phase boundaries a marked pseudo-transition manifested by vigorous
narrow peaks of the specific heat and magnetic susceptibility besides
a sudden change of the entropy and magnetization.

\subsection*{Acknowledgments}

This work was partially supported by Brazilian Agency CNPq and FAPEMIG.

\end{document}